# Reversible Excitonic Charge State Conversion and High Quasiparticle Densities in PVA-doped Monolayer WS$_2$ on 2D Microsphere Array

*Debasish Biswasray, Yogendra Singh[†], Amar Jyoti Biswal[†], and Bala Murali Krishna Mariserla**

Ultrafast Physics Group, Department of Physics, Indian Institute of Technology Jodhpur, Rajasthan- 342030, India

*Corresponding Author Email: bmkrishna@iitj.ac.in, baluuoh@gmail.com

[†] Equal author contribution

**Abstract:**

Controllable quasiparticle radiation in two-dimensional (2D) semiconductors is essential for efficient carrier recombination, tunable emission, and modulation of valley polarization which are strongly determined by both the density and nature of underlying excitonic species. Conventional chemical doping techniques, however, often hinder the reversibility and density of excitonic charge states (exciton and trion) due to unfavorable interactions between dopant and 2D materials. In this work, efficient excitonic charge state conversion is achieved by doping monolayer WS$_2$ using water rinsed PVA and the quasiparticle densities are further enhanced by applying high periodic biaxial strain (up to 2.3%) through a 2D silica microsphere array. The method presented here enables nearly 100% reversible trion-to-exciton conversion without the need of electrostatic gating, while delivering thermally stable trions with a large binding energy of ~56 meV and a high free electron density of ~3 × 10$^{13}$ cm$^{-2}$ at room temperature. Strain-induced funneling of the PVA-injected free electrons substantially increases the excitonic quasiparticle densities and boosts the trion emission by 41%. Overall, this approach establishes a versatile platform for excitonic charge state conversion and enhanced quasiparticle density in 2D materials, offering promising opportunities for future optical data storage, quantum-light and display technologies.

## 1. Introduction:

The advent of atomically thin two-dimensional (2D) transition metal dichalcogenides (TMDs) has redefined the landscape of excitonic physics in semiconductors for the next-generation optoelectronic and quantum photonic devices.[1–3] Monolayer TMDs exhibit strong quantum confinement and dramatically reduced dielectric screening, resulting in tightly bound excitons.[4–6] The binding energies of these excitons range from 0.2 to 0.9 eV, nearly two orders of magnitude larger than conventional semiconductors.[7] Such large binding energies facilitate the emergence of rich many-body phenomena including the formation of charged excitons such as trions, which play a pivotal role in the optical and electronic responses of 2D systems.[8,9] The free carrier density in doped/undoped 2D materials is a critical parameter to govern the densities of both neutral excitons and trions. Since trions carry net charge, it enables to tune their emission by external electric and magnetic fields,[10,11] facilitating optically addressable spin-valley locking.[12] Thus, the tailored optical responses of TMDs can be achieved by controlling the densities of exciton and trion along with their interconversion.

The modulation of free carrier densities with electrostatic gating have been demonstrated by varying the gate voltage to enable controlled injection of carriers in semiconductors for reversible excitonic and trionic emission.[10,13] However, this technique is often comprised of fabrication complexity, environmental instability, and susceptibility of inducing defect states. Alternatively, optical methods such as laser irradiation and orbital angular momentum of light have shown the ability to control the densities of excitons and trions.[14–16] In addition, mechanical methods of non-uniform strain through AFM tip[17] and pressure induced through diamond anvil cell[18] have also shown the exciton-trion conversion. However, these methods are impractical for portable/on-chip devices since they need to be integrated with large and sophisticated equipment for controlling the exciton and trion densities. In contrast, chemical doping provides a facile and stable route to modulate carrier densities through controlled charge transfer between molecular dopants and the TMDs. Molecular dopants, such as 2,3,5,6-tetrafluoro-7,7,8,8-tetracyanoquinodimethane ($F_4TCNQ$) and hexaazatriphenylenehexa-carbonitrile (HATCN) accept electrons from the TMD surface and induces *p*-type doping for positive trion formation.[19,20] Conversely, *n*-type dopant (nicotinamide adenine dinucleotide (NADH)) donate electrons to the TMD surface to increase the negative trion density.[19] Since TMDs facilitate high surface-to-volume ratio, they can efficiently donate/accept electrons to/from physisorbed molecules such as anisole and benzene derivatives that have been reported to change the carrier densities in monolayer TMDs through induced dipole charge transfer and π-π interactions.[21,22] Similarly, polymer-based adsorbents such as poly (methyl methacrylate) (PMMA), poly (dimethylsiloxane) (PDMS), and poly (vinyl alcohol) (PVA) induce *n*-type doping to TMD monolayers.[23,24] Yet these chemical doping approaches often show low carrier densities due to improper interaction between dopant and 2D material, resulting in small trion binding energies which triggers thermal quenching of trion emission at higher temperatures. Furthermore, the excitonic charge state reversibility i.e., the switching from excitons to trions and vice versa with chemical doping have been a quest.

Here, we studied excitonic quasiparticle emission in monolayer $WS_2$ for characterizing charge-state-dependent optical properties. The limitations of conventional doping techniques were addressed by doping monolayer $WS_2$ with PVA and improved interactions through water rinsing process to induce high free carrier density. The reversibility of excitonic charge states were demonstrated to show the trion-to-exciton switching without electrostatic gating. With external stimuli (temperature, laser excitation), we achieved controlled exciton-trion interconversion with stable quasiparticle formation and tunable emission. In addition, we introduced a fabrication-free PVA-assisted transfer method to induce periodic strain in monolayer $WS_2$ through silica microspheres to enhance quasiparticle densities.

## 2. Results and Discussion:

The 2D $WS_2$ crystals were prepared through standard scotch tape exfoliation technique[25] and the number of layers in the crystals were identified through photoluminescence (PL) and Raman spectroscopy. The PL peak around 2.01 eV (617 nm) corresponding to the A-exciton emission and the characteristic frequency difference of ~66 cm$^{-1}$ between the in-plane $E_{2g}^1$ and the out-of-plane $A_{1g}$ vibrational Raman modes confirm the monolayer $WS_2$.[26] The free carrier injection to create quasiparticle charge states (exciton and trion) was achieved by doping the monolayer $WS_2$ with PVA. In brief, 5 wt% of a PVA film was prepared and transferred on to the pristine monolayer $WS_2$. Despite the presence of electron-donating hydroxyl groups in PVA-film, the PL spectrum of the PVA-stacked monolayer clearly showed dominant exciton emission and very weak trion emission. The weak trion emission is due to van der Waal (vdW) gap between PVA film and $WS_2$ monolayer as shown in the top panel of Figure 1a, which inhibits the charge transfer. Through a water rinsing process, the bonding forces [27] between ─OH groups of PVA film and the top layer sulphur atoms of $WS_2$ were established, as illustrated in bottom panel of Figure 1a, to mitigate the charge transfer. The optical image of the prepared PVA-doped monolayer $WS_2$ sample after rinsing for a few minutes is shown in Figure 1b. The PL spectrum of PVA-doped monolayer $WS_2$ along with pristine $WS_2$ are shown in Figure 1c, that reveals two distinct emission peaks at 2.01 eV and ~1.97 eV, corresponding to excitons ($X^0$) and negative trions ($X^-$), respectively. The emergence of trion peak is the direct evidence of successful electron transfer from PVA to WS$_2$ which is established by bonding forces through water rinsing. This causes increased free electron density in monolayer WS$_2$ and renormalization of conduction band. The resulting *n*-type doping through charge transfer was further confirmed by a Raman shift [23,28] of 4 cm$^{-1}$ for the $A_{1g}$ out-of-plane phonon mode, illustrated in Figure 1d. On the other hand, the $E_{2g}^1$ peak showed no shift in the presence of PVA, suggests that the doping process have not introduced any structural distortion.[23,29]

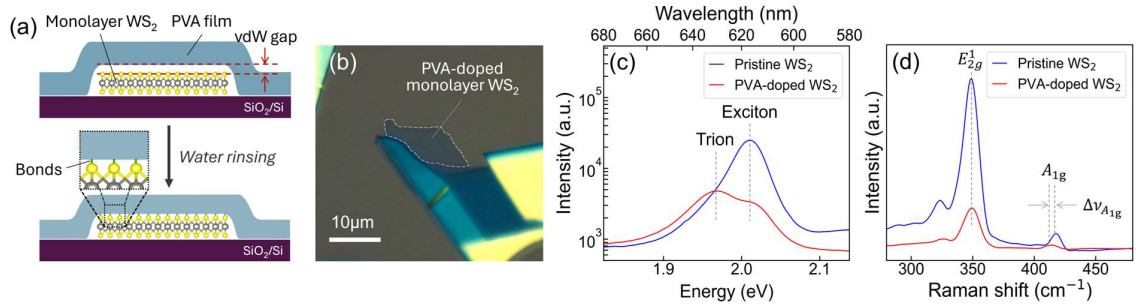

**Figure 1:** (a) Schematic illustration for PVA-doping of monolayer WS$_2$. The top panel illustrates the stacking of PVA-film on monolayer WS$_2$ along with the vdW gap between them. The bottom panel depicts efficient PVA doping achieved after water rinsing process and the magnified section shows the bonds between PVA and WS$_2$. (b) Optical image of the PVA-doped monolayer WS$_2$. (c) Photoluminescence and (d) Raman spectra of pristine and PVA-doped monolayer WS$_2$.

The exciton-trion conversion and evolution of excitonic quasiparticle densities were studied with change in temperature and excitation laser power using a temperature controlled micro-PL system, depicted in Figure 2a. Both exciton and trion emission energies exhibited a pronounced redshift, when the temperature was increased from 80 K to 360 K, as shown in Figure 2c. This energy shift is due to strong electron-phonon interactions at higher temperatures, resulting in band structure renormalization.[5] The temperature dependent evolution of energies ($E_g(T)$) were fitted using the O'Donnell equation, [30]

$$E_g(T) = E_g(0) - S\langle\hbar\omega\rangle[coth(\langle\hbar\omega\rangle / 2k_BT) - 1], \qquad (1)$$

where $\hbar\omega$ represents the phonon energy and $k_B$ is the Boltzmann constant. The electron-phonon coupling interaction strength $S$ was extracted from the fitting. $S$ was higher for trion compared to exciton, which indicates that trion strongly couple to phonon due to additional electron. At higher temperatures, trions dissociate into neutral excitons and free electrons, resulting in exciton dominated PL spectrum, whereas at low temperatures, specifically T < 200 K, the PL spectrum shows strong trion emission. At low temperatures, the injected electrons in *n*-doped monolayer WS$_2$ occupy states near the conduction band minimum at the K valley[28,31,32] to facilitate trion formation, while Pauli blocking limits the increase of exciton density. As temperature increases, the effective mass of electron increases by 4-10% and the band curvature broadens due to strong electron-phonon interaction. This leads to an excitonic band renormalization which relaxes the Pauli blocking and increases the carrier density around K valley, as illustrated in Figure 2b. Hence, the exciton formation becomes more favourable over trions at higher temperatures. This temperature dependent exciton-trion conversion was quantified with trion-to-exciton peak intensity ratio ($I_{X^-}/I_{X^0}$) to signify the relative changes in quasiparticle densities. The $I_{X^-}/I_{X^0}$ decreases exponentially with increase in temperature, which indicates a thermally activated decay in trion density. The change in relative quasiparticles densities with temperature was calculated using an empirical relation, adapted from Kesarwani et al.[15] given by

$$I_{X^-}/I_{X^0} = \gamma_0 e^{-\zeta T}, \qquad (2)$$

where $\gamma_0$ is a fitting constant, and $\zeta$ represents the decreasing trion density along with the corresponding increase in exciton density as a function of temperature ($T$). The obtained low $\zeta$ of ~0.03 K$^{-1}$ signifies that the induced *n*-type doping is sufficiently strong to form thermally stable trions, leading to robust trion emission from PVA-doped monolayer WS$_2$. The spectral weights of excitons and trions (expressed as $\frac{I_{X^0}}{I_{X^0}+I_{X^-}}$ and $\frac{I_{X^-}}{I_{X^0}+I_{X^-}}$, respectively) are plotted as a function of temperature in Figure 2d, which reveals that the trion emission dominates for $T < 320$ K, while the exciton emission dominates for $T \geq 320$ K in the PVA-doped monolayer WS$_2$.

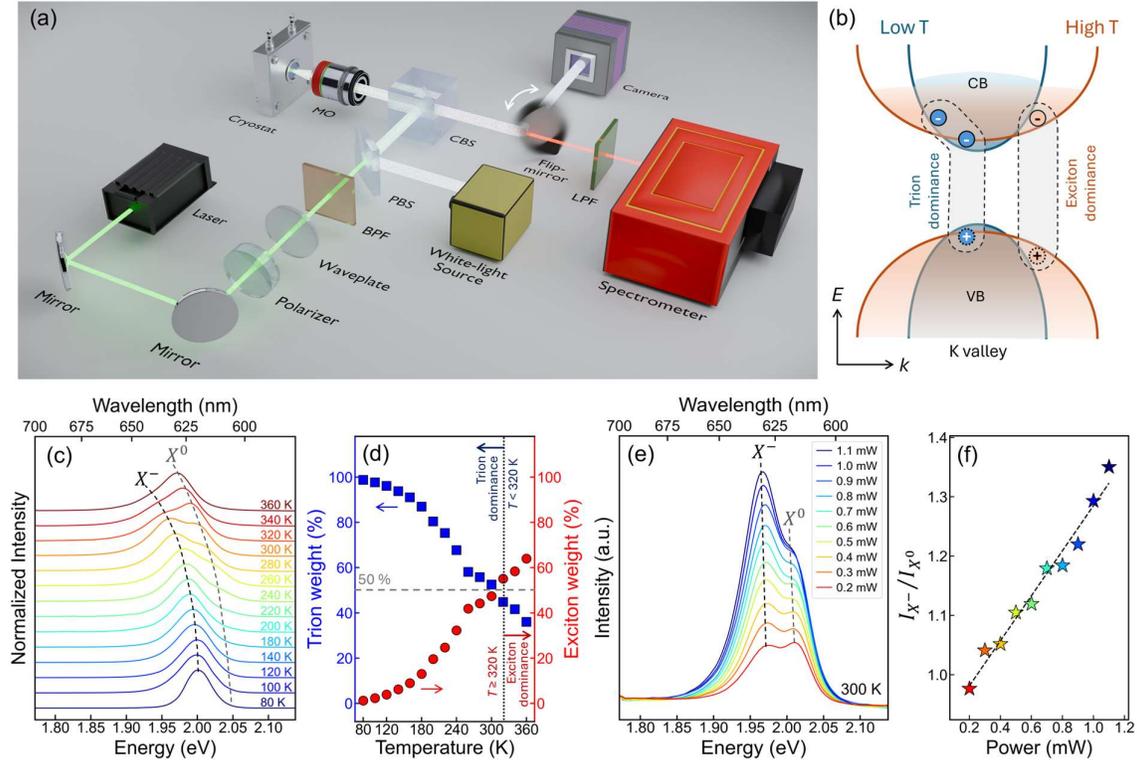

**Figure 2:** (a) Schematic for temperature controlled micro-PL setup (MO: Microscope objective, CBS: Cube beamsplitter, PBS: Plate beamsplitter, BPF: Band-pass filter, LPF: Long-pass filter). (b) Dispersion of *n*-doped monolayer WS$_2$ at K valley at different temperature regimes (VB: valance band and CB: conduction band). (c) Temperature dependent PL spectra and (d) the corresponding spectral weights of excitons and trions, where the horizontal dashed line represents the equal exciton and trion spectral weight, and the vertical dotted line denotes the temperature of transition between the exciton and trion dominance regions, $T < 320$ K and $T \geq 320$ K, respectively. (e) Power dependent PL spectra and (f) the corresponding trion-to-exciton intensity ratio of the PVA-doped monolayer WS$_2$, where the dashed line represents linear fit.

The excitonic quasiparticle emission with laser power also shows variation of charge state densities. On increasing laser excitation power from 0.2 mW to 1.1 mW, exciton emission peak remains unchanged, while the trion peak shows a slight redshift as shown in Figure 2e. The corresponding power dependent variation of exciton and trion peak energies are almost negligible due to localized electron-phonon coupling in comparison to temperature dependent changes where strong electron-phonon coupling prevails across the sample. Notably, the peak intensities of both excitons and trions increased significantly with power due to higher photocarrier density and radiative recombination of the quasiparticles. The corresponding trion-to-exciton intensity ratio $I_{X^-}/I_{X^0}$ is plotted in Figure 2f, which exhibits a linear transition from excitonic to trionic states above 0.2 mW excitation power. At lower excitation power (around 0.2 mW), the ratio is less than 1, suggesting that the photogenerated carriers were insufficient to support significant trion formation. Beyond 0.2 mW, the intensity ratio exceeded unity and increased almost linearly with power, which indicates that the trion formation was favourable with available photocarriers. These power dependent modulations of quasiparticle charge states signifies its critical role in tuning the excitonic densities.

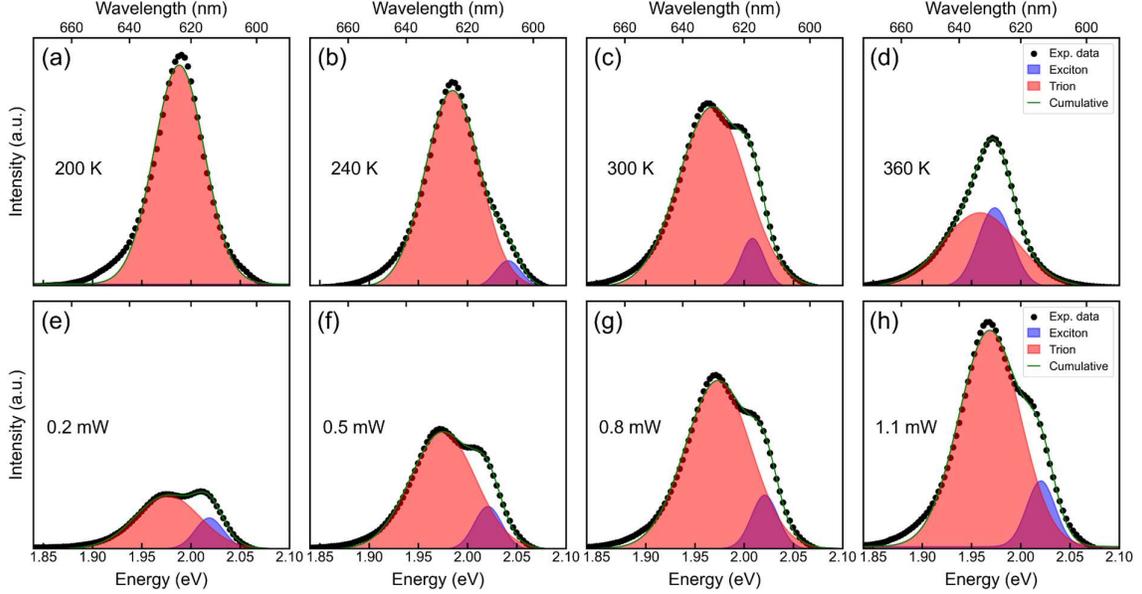

**Figure 3:** Fitting to the PL spectra of PVA-doped monolayer $WS_2$ (a)-(d) at different temperatures, and (e)-(h) at different excitation laser powers.

The spectral characteristics of excitons and trions were extracted by fitting the PL spectra at different temperatures and laser powers, shown in Figure 3. These fittings demonstrate that the relative trion density decreases with increase in temperature due to their thermal dissociation, whereas trion density increases with laser power owing to the enhanced photo-induced carrier generation. The full width at half maxima (FWHM) of the exciton and trion emission, shown in Figure 4a, were increased by ~17 meV and ~24 meV, respectively with temperature raise from 200 K to 360 K. The enhanced broadening in the trion emission indicates a higher sensitivity to thermal perturbations due to the presence of excess carriers. The electrons interaction with phonons led to non-radiative recombination and increased the FWHM for trions compared to excitons.[33] In contrast, power dependent PL spectra showed minimal change in FWHM, as shown in Figure 4b. The FWHM of excitons and trions increased by ~1.97 meV and ~4.86 meV, respectively, over the laser excitation power range (0.2 mW - 1.1 mW). These small FWHM variations indicate weak many-body interactions[34] and negligible power-induced spectral broadening, which can be attributed to the low intrinsic defect density of the exfoliated monolayer $WS_2$. However, the absolute linewidths of trions remained consistently broader than those of excitons across all temperatures and excitation powers. This intrinsic broadening arises from additional non-radiative decay channels associated with trions, such as Auger-like recombination and phonon-assisted scattering.[34–36] These processes act as additional dephasing pathways which shortens the excitonic coherence time and lead to the emission spectral broadening,[34] even in high-purity systems.

The distinct optical response of the neutral excitons and negative trions with respect to temperature and laser power is fundamentally governed by the quantification of trion density and the availability of free electrons near the Fermi-level. The formation of trion is primarily decided by the Coulomb interaction strength between a neutral exciton and free electron, which is quantified as trion binding energy.[37] Hence, the trion binding energy ($E_{X^-}^b$) depends on the energies corresponding to trion ($E_{X^-}$), exciton ($E_{X^0}$), and the Fermi-level ($E_F$), which are related as [10]

$$E_{X^-}^b + E_F = E_{X^0} - E_{X^-}. \qquad (3)$$

Here, the parameter $E_{X^0} - E_{X^-}$ defines the ionization potential of a trion, i.e., the minimum energy required to dissociate an electron from its bound state. Thus, the density of free electrons is an important parameter to quantify the trion formation, which is often correlated with the doped states at the Fermi-level. Theoretically, the free electron density ($n_{e^-}$) in 2D materials at a temperature $T$ can be measured using mass action law,[38]

$$\frac{n_{X^0} n_{e^-}}{n_{X^-}} = \frac{4 m_{eff} k_B T}{\pi \hbar^2} exp\left(-\frac{E_{X^-}^b}{k_B T}\right) \quad (4)$$

where $n_{X^0}$ and $n_{X^-}$ are the densities of neutral excitons and negative trions at the Fermi-level, respectively; and $\hbar$ is the reduced Plank's constant. $m_{eff} = \frac{m_{X^0} m_e^*}{m_{X^-}}$ is the reduced effective mass of monolayer WS$_2$,[39] where $m_{X^0} = \frac{m_e^* m_h^*}{m_e^* + m_h^*}$ is the exciton mass, $m_{X^-} = \frac{m_e^*(m_e^* + m_h^*)}{2 m_e^* + m_h^*}$ is the trion mass, $m_e^* = 0.31\ m_e$ [15,40] is the effective electron mass, and $m_h^* = 0.42\ m_e$ [15] is the effective hole mass ($m_e = 9.11 \times 10^{-31} kg$). The densities of excitons and trions are extracted from the area under curves of trion and exciton emission spectra. [8,37]

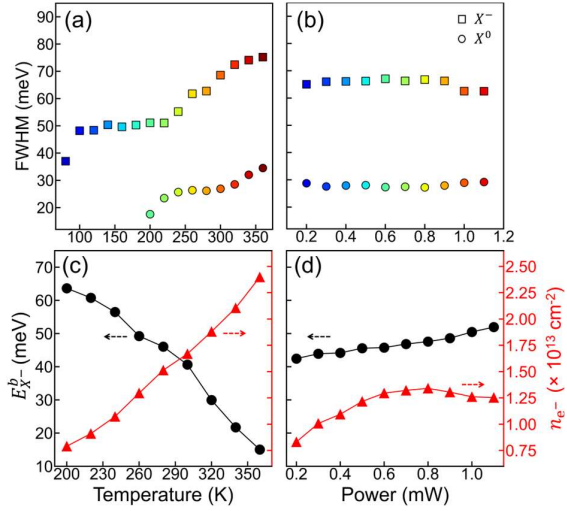

**Figure 4:** Evolution of FWHM of the trion and exciton peaks with respect to (a) temperature and (b) excitation laser power. The filled squares indicate FWHM of trion peaks and the filled circles indicate FWHM of exciton peaks. Variation in trion binding energies and free electron density with respect to (c) temperature and (d) power of laser power.

By solving equations (3) and (4), we determined the trion binding energies and free electron densities as a function of temperature and laser power. With rise in temperature, trion binding energy decreases while increase in the free electron densities shown in Figure 4c ascribed to the dissociation of trions. Conversely, increasing excitation laser power leads to the enhancement of trion binding energy (Figure 4d) resulting in creation of stable trions due to the photo-generated free electrons and optically induced trap states.[41] The room temperature trion binding energy of PVA-doped monolayer WS$_2$ reaches up to 48 meV and the free electron density increases up to ~1.5 × 10$^{13}$ cm$^{-2}$, which is two orders higher than pristine monolayer WS$_2$ at room temperature. Moreover, our obtained free electron density is significantly higher than earlier reported values which is compared and displayed in Table 1.

The excitonic charge state reversibility was investigated by removing the dopant layer of PVA through a simple and continuous water rinsing process (undoping method) to PVA/WS$_2$ sample. Remarkably, the PL emission of water rinsed sample demonstrated an excellent resemblance with the pristine monolayer WS$_2$ and confirms the restoration of pure exciton emission peak shown in Figure 5a, which also provides the evidence that the rinsing process did not introduce any structural modifications or defects. The charge state reversibility was quantitatively analyzed by the trion-to-exciton density ratio ($n_{X^-}/n_{X^0}$) and intensity ratio ($I_{X^-}/I_{X^0}$) for pristine, PVA-doped, and PVA-removed monolayer $WS_2$ samples. These ratios serve as key indicators of the charge state conversion, whose values > 1 signifies trion dominance and < 1 indicates exciton dominance. The Figure 5b shows trion-to-exciton density ratio decreases from ~5 for PVA-doped $WS_2$ to ~0.23 for PVA-removed $WS_2$, which is equivalent to the ratio for pristine $WS_2$. This undoping method reduces the trion emission and restores the neutral exciton emission. When the sample is re-doped with PVA, the intensity ratio increases beyond 1 and after rinsing again, it returns to ~0.23, demonstrating the reversibility of excitonic charge states with doping and undoping cycles. The reversibility was also projected through the trion-to-exciton intensity ratio $I_{X^-}/I_{X^0}$. The doping/undoping cycles with respect to the density and intensity ratios demonstrate the switching between excitonic charge states with high reversibility and reproducibility. Such behavior resembles the electrostatic gating effect, which is not achieved so far in chemical doping of 2D materials.

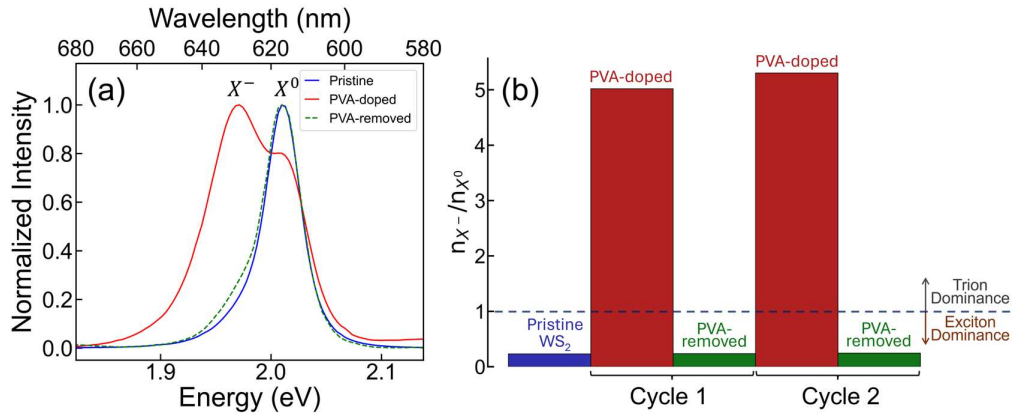

**Figure 5:** (a) PL spectra of doped and undoped $WS_2$ through water rinsing (PVA-removed). (b) trion-to-exciton density ratio for the pristine, PVA-doped, and PVA-removed (undoped) monolayer $WS_2$ samples, illustrating complete reversible doping/undoping cycles. A ratio > 1 indicates trion dominance, while < 1 denotes exciton dominance.

Further, to increase the excitonic charge state densities we introduced periodic localized biaxial strain on monolayer $WS_2$. In order to provide the localized biaxial strain, we developed a PVA-assisted transfer process for suspending $WS_2$ monolayer on a 2D hexagonally close-packed monodispersed silica microspheres, illustrated in Figure 6a. In brief, a PVA-film was stamped onto a mechanically exfoliated monolayer $WS_2$ and peeled off along with the $WS_2$ flake while heating up to 50 $^0C$ for 2 minutes. The PVA/$WS_2$ stack is then transferred on to silica microspheres which were synthesized through a modified Stöber method[42] with an average particle diameter of ~1 μm, as shown in Figure 6b. The elemental composition of the microspheres was identified by energy-dispersive X-ray spectroscopy (EDS), which shows the presence of silicon and oxygen as the most prominent constituents. The PVA/$WS_2$ stack was suspended such that approximately half of the $WS_2$ flake extended beyond the microsphere region (Figure 6c). Thus, it created two regions of interest for the strained monolayer $WS_2$ suspended on 2D microsphere array, and the unstrained monolayer $WS_2$ lying outside microspheres. The PVA-film was then rinsed with water for few minutes to facilitate electron doping into monolayer $WS_2$ across both regions. The PL spectra of the strained and unstrained regions exhibit both exciton and trion emission, those were highly stable even after a couple of months, which indicates that PVA serves as a doping medium as well as a protective layer for $WS_2$ against moisture and surface oxidation.

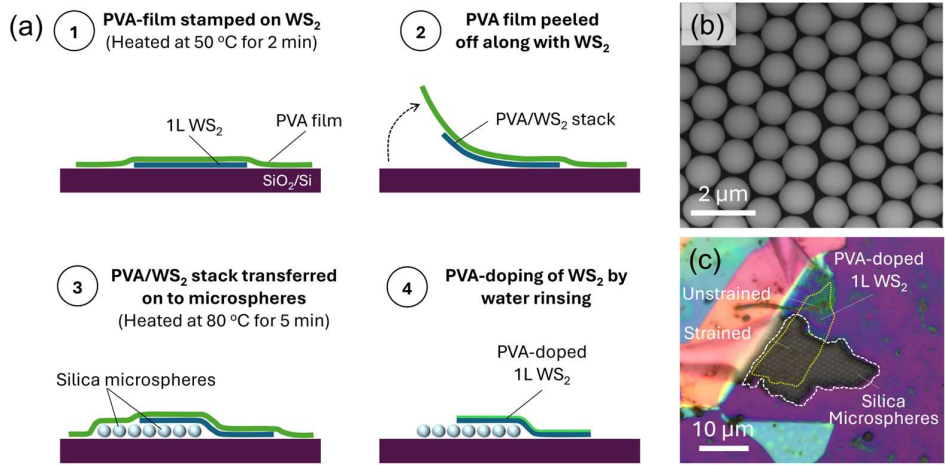

**Figure 6:** (a) Schematic for introducing biaxial strain through PVA-assisted transfer of monolayer (1L) WS$_2$ on silica microsphere array. (b) FE-SEM image of 2D hexagonal monodispersed silica microsphere array. (c) Optical image of PVA-doped monolayer WS$_2$ suspended on silica microsphere array showing two regions of interest (strained and unstrained WS$_2$).

The sagging of the WS$_2$ flake at the interstitial positions of the microspheres governs the strain which can be varied by different sagging depths, as shown in Figure 7a and the size of the microspheres along with their periodic arrangement. We estimated the strain on our samples from the exciton emission energy shift of the strained region compared to the unstrained region, using the relation, $\Delta E = 50\varepsilon$,[29] where $\Delta E$ is the exciton energy shift in meV per strain percentage ($\varepsilon$). The observed exciton emission energies were redshifted, as shown in Figure 7b, by minimum value of 9.5 meV and maximum of 45 meV, corresponding to 0.2% and 2.3% of strain, respectively. The red-shift is ascribed to the strain-induced lattice expansion of doped WS$_2$ and subsequent renormalization of bandgap in conjuction with lowering the conduction band edge.[43,44] It is prominent to achieve a 2.3% of strain on monolayer WS$_2$ through simple PVA-assisted transfer route that is comparable to the maximum sustainable strain of 2.8%, achieved by applying mechanical fore through AFM tip.[17] Such high strain percentage significantly enhanced the trion emission up to 41% as shown in Figure 7b due to the funnelling of free electrons to maximum strain position,[17] i.e, the apex of silica microspheres in our case. These funnelled free electrons at the apex of microspheres bind with photo-generated excitons and create localized trions. However, the 0.2% strain did not show significant increase in trion emission, due to negligible electron funnelling.

The PL spectra of 0.2% and 2.3% strained PVA-doped monolayer WS$_2$ samples were modelled using the framework proposed by Harats et al.[17] In this approach, the spatially averaged PL from the strained region is expressed as

$$\langle PL \rangle = \int \left[ PL_{X^0}(u_{X^0}(r))n_{X^0}(r) + PL_{X^-}(u_{X^-}(r))n_{X^-}(r) \right] r dr \quad (5)$$

where $PL_{X^0}$ and $PL_{X^-}$ are the exciton and trion spectra, respectively, obtained by fitting the PL spectrum of the unstrained region. The functions $PL_{X^0}(u_{X^0}(r))$ and $PL_{X^-}(u_{X^-}(r))$ represent the exciton and trion spectra at zero strain, respectively, shifted by the change in energy ($\Delta E$) induced by local strain. The quantities, $u_{X^0}(r)$ and $u_{X^-}(r)$ denote the local energies of exciton and trion, while $n_{X^0}(r)$ and $n_{X^-}(r)$ denote the exciton and trion densities, respectively. The experimentally measured PL spectra were modelled using Eq. (5) and found a good agreement, as shown in Figure 7b. When strain increased from 0.2% to 2.3%, the trion-to-exciton density ratio also increased significantly, as shown in the top panel of Figure 7c. The enhanced trion emission originated from free electron funnelling, rather than exciton funnelling, which is consistent with the observations of Harats et al.[17] With the aid of electron funnelling effect, strained samples exhibited higher trion binding energy up to ~56 meV and larger free electron density of ~$3 \times 10^{13}$ cm$^{-2}$. Also, the strained sample showed strong increase in trion binding energy as well as free electron density with respect to the input laser power due to the funnelling effect. We obtained a large variation in trion binding energy (~28 meV) for the strained sample in comparison with previous report where it was limited to a maximum of < 10 meV, because of unstable trions at high laser powers.[15] The trion density for lower strain, such as 0% and 0.2% increased almost linearly with respect to the excitation laser power as shown in bottom panel of Figure 7c, due to the rise in photogenerated carrier density. For higher strain of 2.3%, the trion density increased as square root function of power, which is limited by the saturation of electron funnelling on photoexcitation. Our results demonstrate that strain induced through the PVA-assisted transfer method offers the simplest and robust pathway to control exciton and trion densities in monolayer WS$_2$ with size of microspheres and their periodicity.

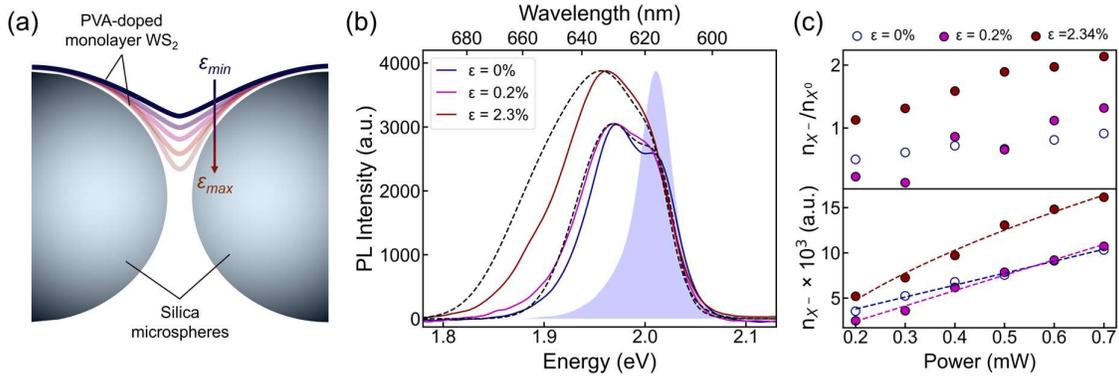

**Figure 7:** (a) Sagging profiles of PVA-doped monolayer WS$_2$ lying between two silica microspheres. The minimum or maximum strain ($\varepsilon_{min}$ or $\varepsilon_{max}$) at the apex of the microsphere corresponds to the minimum or maximum sagging depths, respectively. (b) PL spectra from 0%, 0.2%, and 2.3% strained PVA-doped monolayer WS$_2$, where the black dashed lines represent the model with the spatially averaged PL. The blue shaded area represents the PL spectrum of pristine monolayer WS$_2$, showing the pure exciton peak. (c) Power dependent trion-to-exciton density ratios (top panel) and trion densities (bottom panel) of samples with different strain percentages. Dashed lines represent the linear and square root fit.

Here, we compare our developed method (PVA doping with and without the strain) with previously reported approaches including electrical, optical, and chemical doping in Table 1. Our approach exhibits an exceptionally large trion binding energies up to 48 meV (without strain) and 56 meV (with strain) due to enhanced Coulomb interactions and higher thermal stability of trions. Furthermore, we have obtained high free carrier densities of ~$1.5 \times 10^{13}$ cm$^{-2}$ (without strain) and ~$3 \times 10^{13}$ cm$^{-2}$ (with strain) at room-temperature, which are at least two orders of magnitude higher than pristine monolayer WS$_2$. Our method also demonstrates the excitonic charge state reversibility through simple water rinsing process, without involvement of electric gating. Gaur et al. and Vega-Mayoral et al. have shown trion binding energies up to 48 meV through water-based physisorption on CVD-grown WS$_2$,[28] and liquid phase exfoliated WS$_2$,[24] leveraging the strong dipole moment of water molecules for efficient charge transfer. However, their approach yields low carrier densities of ~$10^{11}$ cm$^{-2}$. Alternatively, Carmiggelt et al. have reported aromatic solvent doping and anisole treatment but the trion binding energy of ~23 meV and a lower carrier density ($\leq 10^{12}$ cm$^{-2}$) were lower.[21] Further, their samples may suffer from environmental instability due to the volatile nature of anisole. While polymer-based dopants such as PMMA and PDMS provide good environmental stability, they typically result in lower trion binding energies.[23] A binding energy of 48 meV was reported with PVA doping to liquid-phase exfoliated (LPE) WS$_2$ flakes,[24] the synthesis process introduces high defect densities with no excitonic charge state reversibility. Optical methods such as change in orbital angular momentum (OAM) of light and laser irradiation,[14,16] achieve a lower trion binding energies (36 and 40 meV) with a carrier density of $7×10^{12}$ cm$^{-2}$.[15] Other methods, such as doping with HATCN[20] and graphene[46] have shown trion binding energies in the range of 26–50 meV and carrier densities of ∼$10^{12}$ cm$^{-2}$, but their charge state reversibility has remained unexplored. Although electrostatic gating on MoSe$_2$ [8] and MoS$_2$ [10] provides excellent charge state reversibility, they are often limited to lower carrier densities ($10^{10}$–$10^{13}$ cm$^{-2}$) and modest trion binding energies (20–30 meV). Other techniques involving F$_4$TCNQ functionalization[19] and plasma treatments[45,47] deliver high carrier densities ($10^{12}$–$10^{13}$ cm$^{-2}$) but lower trion binding energies with lack of charge state reversibility. In comparison to the reported methods, our PVA-based technique shows strong trion binding energies and higher free carrier densities with charge state reversibility. Thus, our PVA-assisted doping technique with and without strain is highly efficient to tailor the excitonic properties for next-generation optoelectronic devices.

**Table 1:** Comparison of different methods for excitonic quasiparticle modulation in 2D materials, based on room temperature trion binding energies ($E_b^{X^-}$), free electron densities ($n_{e^-}$), and charge state reversibility ($X^0 \leftrightarrow X^-$). Various synthesis processes included for this comparison are- mechanical exfoliation (ME), chemical vapour deposition (CVD), and liquid-phase exfoliation (LPE).

| Sl. No. | TMD Material | Synthesis process | Method | $\Delta E_b^{X^-}$ (meV) | $n_{e^-}$ (cm$^{-2}$) | Charge state Reversibility | References |
|---|---|---|---|---|---|---|---|
| 1 | WS$_2$ | ME | PVA<br>PVA and Strain | 48<br>56 | $1.5 \times 10^{13}$<br>$3 \times 10^{13}$ | High | Our work |
| 4 | WS$_2$ | CVD | Water | 48 | ∼$10^{11}$ | -- | [28] |
| 5 | WS$_2$ | ME | Anisole | 23 | -- | -- | [21] |
| 6 | WS$_2$ | CVD | PMMA<br>PDMS | 26<br>30 | -- | -- | [23] |
| 7 | WS$_2$ | LPE | PVA | 48 | -- | -- | [24] |
| 8 | WS$_2$ | CVD | OAM of light | 36 | $7 \times 10^{12}$ | -- | [15] |
| 9 | WS$_2$ | ME | HATCN | 26 | ∼$6 \times 10^{12}$ | -- | [20] |
| 10 | WS$_2$ | CVD | Laser irradiation | 40 | -- | -- | [14] |
| 11 | WS$_2$ | ME | F plasma | ∼32 | ∼$2.5 \times 10^{13}$ | -- | [45] |
| 12 | WS$_2$ | CVD | Graphene | 50 | ∼$10^{12}$ | -- | [46] |
| 13 | WS$_2$ | ME | Pressure | -- | -- | Low | [18] |
| 14 | MoSe$_2$ | ME | Pulsed laser irradiation | 35 | $1.3 \times 10^{11}$ | -- | [16] |
| 15 | MoSe$_2$ | ME | Electrostatic gating | 30 | ∼$10^{10}$ | High | [8] |
| 16 | MoS$_2$ | ME | Electrostatic gating | ∼20 | ∼$10^{13}$ | High | [10] |
| 17 | MoS$_2$ | ME | F$_4$TCNQ | - | ∼$10^{12}$ | -- | [19] |
| 18 | MoS$_2$ | ME | Cl & H Plasma | 28 | ∼$10^{13}$ | -- | [47] |

## 3. Conclusion:

We demonstrated an efficient strategy to modulate the excitonic quasiparticles of monolayer $WS_2$ through combined PVA doping with strain engineering. The hydroxyl-rich PVA offered efficient *n*-type doping to monolayer $WS_2$ through water rinsing process, resulting in stable trion formation with a remarkably large binding energies up to 56 meV and high free electron density of $\sim 3 \times 10^{13}$ cm$^{-2}$ at room temperature. The excitonic charge state reversibility from trion to exciton and vice versa was demonstrated with a simple PVA-removal process through rinsing which offers efficient quasiparticle switching without electrostatic gating. The quasiparticle densities were significantly enhanced through strain induced free electron funnelling, which led up to a 41% increment in trion emission. The laser power dependent study revealed that the trion density increases as a square root function due to the funnelling saturation at higher power and strain. Moreover, excitonic emission from the PVA-doped sample demonstrated great environmental stability due to PVA encapsulation. Our results establish a robust chemical doping approach for achieving highly stable and reversible excitonic species in TMDs, which offers a promising pathway for tunable excitonics for futuristic optoelectronic devices.

## Acknowledgements
DB and YS would like to thank MoE, Govt. of India for financial support through senior/junior research fellowship. M.B.M.K. acknowledges ANRF, India for funding through CRG/2022/008749 and IIT Jodhpur for funding through seed grant I/SEED/BMK/20230017. We acknowledge CRF and AIoT Fab Facility at IIT Jodhpur for giving access to Raman spectrometer and FE-SEM.

## Authors' Contribution
DB, YS, and AJB prepared the $WS_2$ samples through mechanical exfoliation. AJB prepared silica microspheres and DB transferred $WS_2$ on to microsphere array. PVA films were prepared by AJB. Raman spectroscopy and temperature dependent PL experiments were done by DB, and the power dependent PL measurements were performed by YS, AJB, and DB. The strain dependent PL measurements are done by DB. The modelling of PL spectra, analysis of all the results, and manuscript draft preparation were done by DB and BMKM, while all authors contributed to the discussions and manuscript preparation. BMKM conceived the idea and supervised the project.

## Data Availability Statement
All data are available in the manuscript and in the Supporting Information. Additional information may be made available on request.

## Conflicts of interest
The authors declare no competing financial interest.